\documentclass[12pt,article,nofootinbib]{revtex4}
\usepackage{pdfpages}
\usepackage{epsf}
\usepackage{subfigure}
\usepackage{amsmath}
\usepackage{eqnarray}
\usepackage{epstopdf}
\usepackage{mathrsfs}
\usepackage{natbib}
\usepackage{amssymb}
\usepackage{dcolumn}
\usepackage{graphicx}
\usepackage{color}
\usepackage{hyperref}
\usepackage{enumerate}
\usepackage{float}
\usepackage{tikz}
\usetikzlibrary{decorations.pathmorphing, arrows.meta}

\usepackage{multirow}
\usepackage{blindtext}

\makeatletter
\newcommand*{\rom}[1]{\expandafter\@slowromancap\romannumeral #1@}
\makeatother
\def\be{\begin{equation}}
\def\ee{\end{equation}}
\def\ba{\begin{eqnarray}}
\def\ea{\end{eqnarray}}

\graphicspath{{images/}}

\begin{document}
\title{A Brief Review of Quantum Tunneling:\\
	Computational Approaches and Experimental Evidence}
\author{Sareh Eslamzadeh}
\email{Sareh.Eslamzadeh@gmail.com}
\affiliation{Department of Physics, College of Sciences, Yasouj University, 75918-74934, Yasouj, Iran.}	
\author{Saheb Soroushfar}
\thanks{Corresponding Author}
\email{Soroush@yu.ac.ir}
\affiliation{Department of Physics, College of Sciences, Yasouj University, 75918-74934, Yasouj, Iran.}

\begin{abstract}
This paper presents a concise review of the quantum tunneling approach to Hawking radiation, covering its theoretical foundations, extensions, and experimental efforts. We begin by outlining the Hamilton-Jacobi and Parikh-Wilczek methods, which provide a semi-classical framework for deriving Hawking radiation from stationary black holes. The discussion is then extended to dynamical black holes, where evolving horizons require modified treatments incorporating trapping horizons, Kodama vectors, and dynamical surface gravity. We explored the possible tunneling paths for particles crossing the horizon in dynamical black holes and emphasized the crucial role of the imaginary part of the action in determining the Hawking temperature. In the second part, we review experimental investigations of Hawking radiation, including analogue black hole experiments, quantum simulations, and astrophysical searches for primordial black hole evaporation. While no direct detection of Hawking radiation has been achieved, recent advances in Bose-Einstein condensates, optical analogues, and superconducting qubits offer indirect support for the tunneling interpretation of black hole evaporation.


\end{abstract}

\maketitle
\newpage
\tableofcontents
			
	\section{Introduction}
The fact that light rays do not converge at the event horizon of a black hole, or more precisely, that the area of the event horizon does not decrease, was an early indication of its connection with entropy and the second law of thermodynamics \cite{71Haw} . This insight led to the formulation of laws of black hole thermodynamics, establishing a deep analogy between classical thermodynamics and black hole physics \cite{73Bek, 73Bar}. Building on this foundation, Hawking’s seminal work in quantum field theory in curved spacetime demonstrated that black holes are not entirely black but instead emit thermal radiation, now known as Hawking radiation. By considering a massless scalar quantum field in the background of a gravitationally collapsing black hole, Hawking showed that if a quantum state is initially defined as a vacuum state in the past, it may evolve into a state containing particles in the future. Through the calculation of Bogoliubov coefficients, he deduced that the spectrum of emitted particles follows a black-body distribution with a temperature proportional to the black hole's surface gravity \cite{1974Haw, 1975Haw}. Following Hawking’s discovery, extensive research has been conducted to refine and extend these ideas; as a good example, one can refer to Ref. \cite{76Dam} and the references therein. After this intuitive insight, tunneling calculations were further developed through the work of Gibbons and Hawking on the Euclidean quantum gravity method \cite{76Gib}. Around the same time, the trace anomaly method was employed to compute the expectation value of the energy-momentum tensor components, leading to an alternative derivation of the Hawking flux \cite{77Chri}. More recently, quantum tunneling methods have gained substantial attention as an alternative approach to deriving Hawking radiation. Two major tunneling methods have been developed. First, the Hamilton-Jacobi method, introduced by Srinivasan and Padmanabhan \cite{98Sri}, applies the semi-classical WKB approximation to analyze particle emission. Meanwhile, the null geodesic method, proposed by Parikh and Wilczek \cite{99Parikh, 04Parikh}, interprets Hawking radiation as a quantum mechanical tunneling process through the horizon, while also incorporating back-reaction effects. These tunneling methods provide an intuitive and physically transparent framework for studying black hole radiation and form the foundation of our present research. Another notable method, based on gravitational anomalies and diffeomorphism invariance, was proposed by Robinson and Wilczek \cite{05Rob}, though it falls beyond the scope of this paper.
 
To fully understand the tunneling process, it is crucial to distinguish between stationary and dynamical black holes, as a nature of the horizon significantly affects the emission mechanism and the applicability of semi-classical methods. Stationary black holes, described by time-independent metrics (e.g., Schwarzschild, Kerr), possess well-defined event horizons and constant surface gravity, allowing straightforward applications of quantum field theory for Hawking radiation. In contrast, dynamical black holes, characterized by evolving horizons (e.g., McVittie, Vaidya, and LTB spacetimes), exhibit time-dependent surface gravity and require generalized definitions such as trapping horizons. This distinction complicates the tunneling process, as particle creation and horizon fluctuations influence the emission spectrum, necessitating alternative frameworks beyond traditional semi-classical methods. Indeed, the challenge lies in the fact that key properties of stationary black holes—such as the event horizon, surface gravity, and entropy—are typically defined using universal conditions, such as space-like hypersurfaces and asymptotically flat spacetime. However, for dynamical black holes, these definitions must be reformulated in terms of local concepts. We will examine these quantities in more detail in Section III. Additionally, for a more comprehensive overview of Hawking radiation and its theoretical developments, we refer the reader to Refs. \cite{11Maj, 11Van}.

Hawking radiation is one of the most significant predictions in theoretical physics, connecting quantum mechanics, general relativity, and thermodynamics. Despite its strong theoretical foundation, detecting this radiation directly remains extremely challenging. The main reason is its very low temperature, which is much lower than the Cosmic Microwave Background (CMB), making it practically invisible in current astrophysical observations. Because of this difficulty, researchers have developed alternative ways to study Hawking radiation. These methods can be divided into three main categories: analogue black hole experiments, where event horizons are recreated in laboratory conditions; quantum simulations, which use engineered quantum systems to model black hole physics; and astrophysical observations, particularly searches for radiation from primordial black holes. Each of these approaches provides indirect but valuable insights into the nature of black hole evaporation. In section IV, we will discuss each of these methods in detail and present the related experimental and observational studies conducted so far.

In this paper, we first reviewed tunneling methods applied to the event horizons of stationary black holes in Section II. We then extended the discussion to dynamical black holes in Section III, focusing on the tunneling process from trapping horizons. In Section IV, we explored experimental progress in verifying Hawking radiation. Finally, in Section 5, we summarized the key insights on the tunneling process.

\section{Tunneling from Horizon of the Stationary Black Hole}
It can be argued that a stationary gravitational field can lead to particle creation if the spacetime contains a black hole \cite{98Fro}. In this context, when a pair of virtual particles forms just inside the event horizon, the positive-energy particle can tunnel out, becoming a real particle. Conversely, if a virtual pair is generated slightly outside the horizon, the negative-energy particle can tunnel into the black hole. In both scenarios, the negative-energy particle is absorbed by the black hole, leading to a gradual decrease in its mass, while the positive-energy counterpart escapes to infinity, where it is observed as Hawking radiation. This tunneling effect manifests as shrinking the black hole's horizon radius.
Another important aspect of this process is the role of the particle's action. For a particle falling into the black hole, the action remains real, whereas for an outgoing particle, the action acquires an imaginary component, which governs the tunneling probability. The amplitude of this quantum tunneling process is therefore determined by the imaginary part of the action of the emitted particle. Furthermore, near the event horizon, the angular components of the field equations can be neglected, allowing us to focus on the case of spherically symmetric emission corresponding to the angular quantum number $l=0$, commonly referred to as spherical waves. Consequently, the tunneling method primarily relies on calculating the imaginary part of the action for the emission of spherical waves from the horizon, which is directly linked to the Boltzmann factor and Hawking temperature.\\
With this foundation, we now turn our attention to two widely used mathematical approaches for computing the imaginary part of the action: The null geodesic method, developed by Parikh and Wilczek, which interprets Hawking radiation as a quantum tunneling process incorporating back-reaction effects; The Hamilton-Jacobi method, which formulates the tunneling problem using the semi-classical WKB approximation for scalar field propagation.

\subsection{Parikh-Wilczek Method}
As mentioned earlier, there is no classical trajectory that allows a particle to cross the black hole horizon from the inside. Consequently, virtual particles can escape by tunneling through the horizon in a semi-classical manner. This forms the fundamental premise of the Parikh-Wilczek (P-W) method. In other words, the tunneling probability is determined by calculating the imaginary part of the action for a particle with momentum $p_r$ tunneling from $r_{in}$ to $r_{out}$ , as follows

\begin{equation}\label{IMS1}
\text{Im}\:S = \text{Im} \int_{r_{\text{in}}}^{r_{\text{out}}} p_r \, dr 
= \int_{r_{\text{in}}}^{r_{\text{out}}} \int_0^p dp_r' \, dr 
= \text{Im} \int_{r_{\text{in}}}^{r_{\text{out}}} \int_0^H \frac{dH'}{\dot{r}} \, dr	
\end{equation}

The final term on the right-hand side of the equation is derived using Hamilton’s equations. In the Parikh-Wilczek (P-W) method, another crucial aspect is that the height of the tunneling barrier is determined by the particle itself. This phenomenon, known as the self-gravitation effect, is discussed in Ref. \cite{95Kra}. Their findings indicate that if we assume the total mass remains constant while allowing the black hole's mass to change, a particle with energy $\omega$ follows a geodesic trajectory described by the modified metric, where $M$ is replaced with $M-\omega$. In other words, due to self-gravitational effects, the energy of the emitted particle influences its motion, requiring the replacement of $M$ with $M-\omega$ in the metric. Finally, to properly describe the trajectory of a massless particle, it is essential to use a non-singular metric at the horizon, which is obtained by applying the Painlev\'{e} transformation \cite{21Pain}. Incorporating all these considerations, we can rewrite Eq. \ref{IMS1} as follows

\begin{equation}
	\text{Im}\:S = -\text{Im} \int_{r_{\text{in}}}^{r_{\text{out}}} \frac{dr}{r - r_H} \int_0^\omega \frac{d\omega'}{\kappa [M - \omega']}.
\end{equation}
Where $r_H$ and $\kappa$ are horizon’s radius and surface gravity, respectively. For an evaluation of the integral, we note that it contains a pole at the horizon, requiring the use of residue calculus to solve it. After computing the imaginary part of the action, Parikh and Wilczek applied the WKB approximation to determine the black hole temperature. Within this, by considering the contributions of both particle and antiparticle tunneling, the transmission coefficient, $\Gamma$, for the classically forbidden region is expressed, on one hand, as a function of the imaginary part of the action, and on the other hand, it is related to the emission and absorption probabilities, given by

\begin{equation}\label{Gamma}
	\Gamma = \frac{P_{\text{emit}}}{P_{\text{abs}}} \sim \exp\left( -\frac{2}{\hbar} \, \text{Im}\:S \right) = \exp(-\beta \omega).
\end{equation}

The final term on the right-hand side of the equation is derived from Hawking and Hartle’s work \cite{1976Har}, where $\beta^{-1}$ denotes the black hole temperature.

\subsection{Hamilton-Jacobi Method}
\begin{figure}
	\centering
	\includegraphics[width=0.7\linewidth]{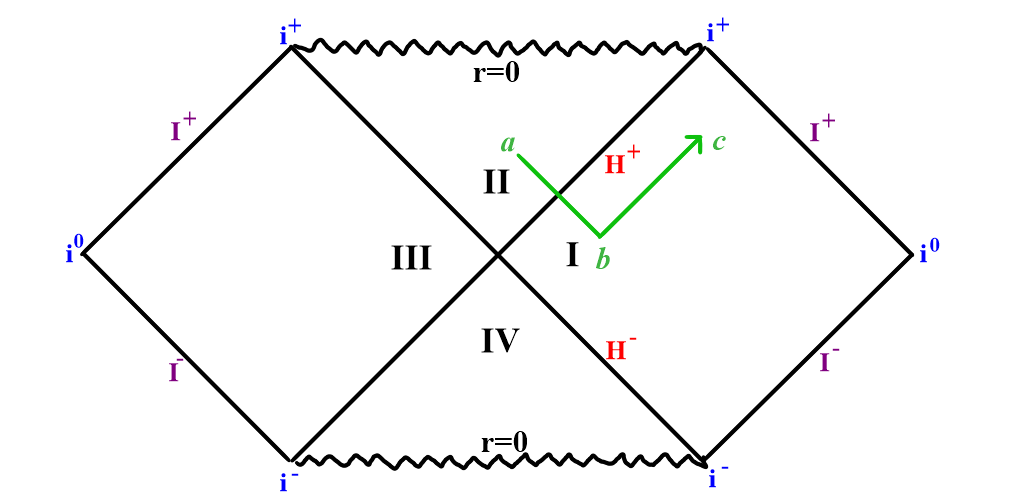}
	\caption{\centering The spherical wave passing through the horizon of Schwarzschild black hole.\newline ($\vec{abc}$) is a light-like geodesic as a continuous path from the inside to the outside of the black hole, where ($\vec{ab}$) is the classical forbidden path and, therefore, in the reverse time direction.}
	\label{Fig1}
\end{figure}

In this method, there is no need to select a specific Painlev\'{e} coordinate, ensuring that the causal structure of Hawking radiation and its back-reaction effects remain preserved. Moreover, this approach extends beyond the assumption of spherical symmetry and can be applied to time-dependent and slowly varying spacetimes. The Hamilton-Jacobi method follows a systematic procedure: it begins with the Klein-Gordon equation for a scalar field propagating in spacetime. The semi-classical wave function of the field is then expanded in terms of the action, assuming that the action of the tunneling particle satisfies the relativistic Hamilton-Jacobi equation. The total action is constructed using its partial derivatives and incorporating the underlying symmetries of the system. Next, the geodesic integral is separated into two distinct parts: The first segment describes the trajectory as the particle crosses the horizon; The second segment extends the motion in the spacetime region outside the horizon, as illustrated in Fig. \ref{Fig1}.
Therefore, the imaginary part of the action can be expressed as follow
\begin{equation}
	\text{Im}\: S = \text{Im} \int_{a \to b \to c} \left(\partial_r S \, dr + \partial_{t_P} S \, dt_P \right). 
\end{equation}
Eventually, by applying near-horizon approximations to the first integral and incorporating Feynman’s $i\epsilon$ rules, the imaginary part of the classical action is derived in terms of the surface gravity and the energy of the tunneling particles. Consequently, the Hawking temperature of the black hole can be determined using Eq. (\ref{Gamma}).\\
Based on the calculations presented, it is crucial to emphasize that the imaginary part of the action is the key factor influencing the Hawking temperature. This is because, in the tunneling approach, an emitted particle must move across a classically forbidden region, moving from inside the horizon to infinity. Due to the causal structure of black hole spacetimes, the event horizon is spatially positioned in the future relative to the external region, meaning that an outgoing particle undergoes time-reversed motion to escape. As a result, its classical action acquires an imaginary component, which directly determines the tunneling probability. Consequently, it is this imaginary contribution, rather than the real part of the action, that governs the thermal spectrum of Hawking radiation and ultimately defines the Hawking temperature.

	\section{Extension of Tunneling Methods to Dynamical Black Hole}
In this section, we aim to extend the semi-classical framework of stationary black holes to dynamical black holes. Unlike their stationary counterparts, dynamical black holes are still characterized by mass, angular momentum, and electric charge, but these parameters evolve over time. This naturally raises a fundamental question: What defines the surface of a dynamical black hole, and where does it form? The challenge lies in the fact that key properties of stationary black holes—such as the event horizon, surface gravity, and entropy—are typically defined using universal conditions, such as spacelike hypersurfaces and asymptotically flat spacetime. However, for dynamical black holes, these definitions must be reformulated in terms of local concepts. To address this, we introduce three essential quantities:
\begin{itemize}
\item{1.} 
Trapping Horizon \cite{08Gour,1994Hay}: A locally defined surface that generalizes the event horizon for evolving spacetimes.

\item{2.} 
Kodama Vector \cite{1979Kodama}: A generalization of the Killing vector field in non-stationary settings, crucial for defining conserved quantities.

\item{3.} 
Dynamical Surface Gravity \cite{1998Hay}: A locally defined measure of surface gravity that accounts for time-dependent horizon evolution.

\end{itemize}
To further clarify these concepts and explore the tunneling process in dynamical black holes, we proceed to derive the temperature of a spherically symmetric dynamical black hole in the following.\\
A spherically symmetric metrics can be locally written as
\begin{equation}
	ds^2 = \gamma_{ij}(x^i) dx^i dx^j + R^2(x^i) d\Omega^2, 
\end{equation}
where $d\Omega^2$ is a normal metric to the sphere of symmetry, as follows
\begin{equation}
	d\gamma^2 = \gamma_{ij}(x^i) dx^i dx^j = -E(r,t) dt^2 + 2F(r,t) dt \, dr + G(r,t) dr^2.
\end{equation}
According to the given metric, the following expression allows us to determine the trapping horizon based on Hayward's definition.
\begin{equation}\label{Xi}
	\chi(t,r) = \gamma^{ij} \partial_i R \partial_j R = \gamma^{rr}(t,r) = \frac{E}{EG + F^2} = 0.
\end{equation}
Besides, using the original definition of the Kodama vector given by $\nabla^\nu (K^\mu G_{\mu\nu}) = 0$, one can derive the Kodama vector for the specified spherically symmetric metric as follows
\begin{equation}
	K^i (x) = \frac{1}{\sqrt{-\gamma}} \, e^{ij} \partial_j R = \left( \frac{1}{\sqrt{EG+F^2}}, 0, 0, 0 \right).
\end{equation}
In fact, Kodama vector is a light-like on the trapping horizon and a space-like inside it. This characteristic suggests that $K^{\mu}$ can be regarded as a generalization of the Killing vector field in stationary black hole spacetimes. Accordingly, it can be expressed as $K^\mu \nabla_{[\nu} K_{\mu]} \approx \kappa K_\nu$, where $\kappa$ represents the dynamical surface gravity, given by
\begin{equation}
	\kappa_H = \frac{1}{2} \left. \Box_{\gamma} r \right|_H = \left. \frac{1}{2F^3} \left( E' F - \frac{1}{2} \dot{E} G \right) \right|_H,
\end{equation}
Where, $\Box_{\gamma}$ is the two-dimensional Klein-Gordon operator, while dot and prime represent differentiation with respect to t and r, respectively. For a more detailed exploration and derivation of these equations, we refer the reader to Ref. \cite{11Van}.

\begin{figure}
	\centering
	\includegraphics[width=0.5\linewidth]{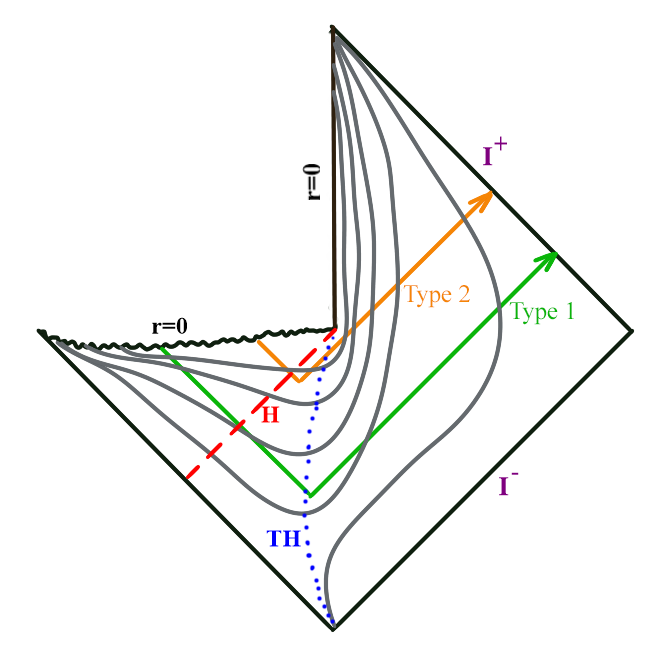}
	\caption{\centering Two types of paths for escaping particles from dynamical black holes.}
	\label{Fig2}
\end{figure}

Before proceeding with the calculations, it is essential to address a fundamental difference in the trajectory of particles in dynamical black holes. By definition, the trapping horizon consists of points where outgoing null rays have zero velocity. As a result, photons remain at rest on the horizon and escape over a characteristic dynamical timescale of $\kappa^{-1}$. Accordingly, in the context of dynamical black holes, there exist two distinct types of tunneling paths, depending on whether the particle is created just outside or inside the trapping horizon, as follows:
	
	\textbf{Type 1 Path:} A pair of particles forms outside the trapping horizon. In this case, the trajectory includes a null ray tunneling backward in time, crossing the trapping horizon from the singularity to its past at infinity.
	
	\textbf{Type 2 Path:} Crucially, type 2 paths are unique to dynamical black holes. These trajectories may be classically allowed, meaning that they do not contribute to the imaginary part of the action and therefore do not affect the tunneling probability.

The Hamilton-Jacobi equation for the tunneling particle is expressed as follows
	\begin{equation}
		\chi (\partial_r S)^2 - \frac{2 \omega F}{\sqrt{EG + F^2}} \partial_r S - \omega^2 G = 0,
	\end{equation}
where $\omega = -\frac{\partial_t S}{\sqrt{EG + F^2}}$ represents the energy of the particle. So, we have
	\begin{equation}
		\partial_r S = \frac{\omega F}{\sqrt{EG + F^2} \, \chi} \left( 2 + O(\chi) \right),
	\end{equation}
where $\chi$ is an expansion of Eq. (\ref{Xi}) near the horizon as follows
	\begin{equation}\label{Approx}
		\chi \approx \dot{\chi} \delta t + \chi' \delta r = \left. \left(\chi' - \frac{G}{2F} \dot{\chi} \right) \right|_H (r - r_H) + \cdots = 2 \kappa (r - r_H) + O((r - r_H)^2).
	\end{equation}
	
The action in terms of the partial derivatives would be as follows
	\begin{equation}
		S = \int_{\gamma} \left( dr \, \partial_r S + dt \, \partial_t S \right) = \int_{\gamma} dr \left[ \partial_r S + \frac{1}{2} G_H \omega_H \right].
	\end{equation}
	
Since there is a pole at $\chi = 0$, applying Eq. (\ref{Approx}) along with Feynman’s rules, the imaginary part of the action is obtained as follows
	\begin{equation}
		\text{Im} S = \text{Im} \int_{\gamma} dr \, \partial_r S = \text{Im} \int_{\gamma} dr \:\frac{\omega F}{\sqrt{EG + F^2} \, \chi}\: \frac{1 + \sqrt{1 + O(\chi)}}{2 \kappa_H (r - r_H - i\epsilon)} = \frac{\pi \omega_H}{\kappa_H}.
	\end{equation}
Finally, we establish that the imaginary part of the action is directly governed by the dynamical surface gravity, highlighting its fundamental role in the tunneling process. This result confirms the connection between black hole thermodynamics and the semi-classical description of particle emission. Moreover, by incorporating the effects of horizon evolution, we show that the temperature of the dynamical black hole can be derived by applying Eq. (\ref{Gamma}), which extends the standard Hawking temperature formula to a time-dependent spacetime.

\section{Experimental Evidence of Hawking Radiation}
Despite the strong theoretical foundation of Hawking radiation, a major challenge remains in its direct observational verification. Due to the fact that a temperature of Hawking radiation from astrophysical black holes is several orders of magnitude lower than the Cosmic Microwave Background (CMB), making direct detection infeasible with current astronomical instruments. Nonetheless, various experimental approaches have been developed to study Hawking-like radiation in controlled laboratory settings and indirect astrophysical searches.
These approaches can be classified into three main categories:\\
\textbf{Analogue Black Hole Experiments:} Laboratory systems that mimic black hole horizons.\\
\textbf{Quantum Simulations:} Engineered quantum systems replicating Hawking radiation properties.\\
\textbf{Astrophysical Observations:} Indirect searches for Hawking radiation from primordial and astrophysical black holes.

\subsection{Analogue Black Hole Experiments}
Analogue gravity experiments provide a way to mimic black hole horizons and study Hawking radiation in controlled settings. These experiments use various physical systems, including fluids, Bose-Einstein Condensates (BECs), optical media, and laser filaments, to create an effective event horizon and analyze the resulting quantum emissions. Among these approaches, BECs have been particularly successful in reproducing Hawking-like radiation. Observations in a black hole laser created in a BEC confirmed the presence of negative-energy partner modes, providing evidence for self-amplified Hawking radiation \cite{14Stein}. Further studies demonstrated direct quantum entanglement between the emitted Hawking-like radiation and its negative-energy counterpart, strengthening the case for the analogy between analogue systems and black hole evaporation \cite{16Stein}. Moreover, a major breakthrough came with the experimental measurement of the Hawking temperature in a BEC-based analogue black hole, where results showed a thermal spectrum matching theoretical predictions \cite{19Stein}. The theoretical framework supporting these findings was established earlier, showing that quantum fluctuations near the event horizon in BECs follow the Bogoliubov dispersion relation \cite{09Recati}.

Another experimental approach involves optical and laser-based analogues. Investigations into ultrashort laser pulse filaments in nonlinear dielectrics suggested the possibility of Hawking-like emission, though this interpretation was later challenged. Alternative explanations proposed that the observed photons could be a result of cosmological particle creation mechanisms rather than genuine Hawking radiation \cite{10Bel,12Unruh}. More recent work has reported experimental signatures of Hawking radiation in high-energy physics experiments, where quantum radiation was analyzed from an accelerated boundary in quantum electrodynamics \cite{24Lynch}.\\
Beyond BECs and optical systems, other analogue models have been developed to simulate black hole phenomena. Studies on slow light propagation in atomic media demonstrated how black hole physics can be explored in optical systems \cite{02Leo}. Additionally, fluid flow experiments have recreated trapping horizons, mimicking the structure of event horizons in real black holes. These systems offer further support for the validity of analogue gravity as a means to investigate quantum effects associated with event horizons.\\
While these experiments provide strong evidence in favor of the semi-classical description of black hole evaporation, ongoing discussions continue regarding alternative interpretations. Further confirmation of stimulated Hawking radiation in analogue black holes has been reported, although challenges remain in observing black hole lasing effects \cite{22Stein}. Some studies suggest that additional mechanisms, such as the Bogoliubov-Cherenkov-Landau (BCL) effect, could influence the observed phenomena, leaving open questions about the exact nature of analogue Hawking radiation.

\subsection{Quantum Simulations of Hawking Radiation}
Beyond analogue black holes, quantum simulations provide an alternative approach to studying Hawking radiation by allowing direct manipulation of quantum tunneling effects and entanglement dynamics. Recent developments in quantum information science have enabled quantum simulations of Hawking radiation in engineered quantum systems, allowing direct observation of quantum tunneling and entanglement across an artificial event horizon. A 2023 experiment using superconducting qubits \cite{23Shi} successfully simulated quantum tunneling and confirmed Hawking-like entanglement entropy growth, further supporting the semi-classical approach to black hole evaporation. In addition to superconducting platforms, research has explored holographic duality, quantum information models, and tensor networks as methods for investigating black hole evaporation and the information loss problem in controlled quantum settings. These approaches contribute to a deeper understanding of black hole thermodynamics and the relationship between quantum mechanics and gravity.

\subsection{Astrophysical Observations of Hawking Radiation}
Although laboratory experiments provide valuable insights into Hawking radiation, astrophysical observations offer the most direct test of this phenomena. If primordial black holes (PBHs) formed in the early universe, their small mass would enable them to emit high-energy photons through Hawking radiation, which could appear as gamma-ray bursts or cosmic rays. Calculations of the expected Hawking radiation spectrum from PBHs indicate that such emissions could be detected in the MeV–GeV energy range using next-generation telescopes like e-ASTROGAM and AMEGO \cite{21Coogan}. Another promising detection method involves multi-messenger observations, where small black holes, or "morsels," created in astrophysical black hole mergers are predicted to produce Hawking radiation in the TeV gamma-ray range \cite{24Cacc}. A potential correlation between gravitational wave signals from detectors such as LIGO and Virgo, besides high-energy photon emissions has been suggested as a way to identify this radiation. While no definitive evidence has been obtained so far, upcoming space-based telescopes and high-energy particle detectors may offer new opportunities to detect its remnants. The search for PBHs evaporation signatures remains a key way in the effort to verify the existence of Hawking radiation.

\section{Summary}
In this paper, we reviewed the quantum tunneling approach to Hawking radiation, emphasizing both its theoretical framework and experimental perspectives. The Hamilton-Jacobi and Parikh-Wilczek methods provide a semi-classical approach to understanding particle emission from stationary black holes, where the imaginary part of the action plays a crucial role in determining the Hawking temperature. We then explored the extension of these methods to dynamical black holes, highlighting the role of trapping horizons, Kodama vectors, and evolving surface gravity in modifying the tunneling process. We clarified  that, in the presence of a trapping horizon for a dynamical black hole, there exist two possible tunneling paths for a particle crossing this horizon. However, what ultimately determines the Hawking temperature is the tunneling of particles created just outside the trapping horizon, as these are classically forbidden from escaping and therefore contribute to the imaginary part of the action, governing the tunneling probability.

In the next section, we examined the major challenge of directly observing Hawking radiation, primarily due to its extremely low temperature compared to the CMB, and explored the experimental evidence that supports these theoretical foundations. At the same time, experimental research in analogue gravity, quantum simulations, and astrophysical searches has provided valuable indirect insights. Bose-Einstein condensate experiments have confirmed key aspects of Hawking-like emission, including entanglement and stimulated radiation, while superconducting qubit simulations have successfully modeled quantum tunneling across engineered event horizons. Furthermore, recent proposals suggest that primordial black holes or black hole merger remnants could emit detectable Hawking radiation, though no conclusive evidence has been found yet. Bridging the gap between theoretical predictions and experimental verification remains a major challenge, but ongoing advancements indicate that the quantum nature of black holes may soon be tested with increasing precision.

\end{document}